\documentclass[preprint,12pt]{elsarticle}

\usepackage{amssymb}

\usepackage{epstopdf}

\journal{Journal of Alloys and compounds}

\begin{document}

\begin{frontmatter}



\title{Role of Sn impurity on electronic topological transitions in 122 Fe-based superconductors}

\author[1,2]{Haranath Ghosh\fnref{email}}
 \author[1]{Smritijit Sen}
\address[1]{Homi Bhabha National Institute, Anushaktinagar, Mumbai 400 094, India.}
\address[2]{Indus Synchrotrons Utilization Division, Raja Ramanna Centre for Advanced Technology,
Indore -452013, India.}
\fntext[email]{Corresponding author : hng@rrcat.gov.in}


\begin{abstract}
We show that only a few percentage of Sn doping at the Ba site on BaFe$_2$As$_2$,  
 can cause electronic topological transition, namely, the Lifshitz transition. 
A hole like d$_{xy}$ band of Fe undergoes electron like transition due to 4\% Sn doping. 
Lifshitz transition is found in BaFe$_2$As$_2$ system around all the high symmetry points. Our detailed first 
principles simulation predicts absence of any Lifshitz transition in other 122 family compounds like 
SrFe$_2$As$_2$, CaFe$_2$As$_2$. This work bears practical significance due to the facts 
that a few percentage of Sn impurity is in-built in tin-flux grown single crystals method
of synthesizing 122 materials and inter-relationship among the Lifshitz transition, magnetism and superconductivity.
\end{abstract}


\end{frontmatter}


\section{Introduction}
The phase diagrams of Fe-based superconductors (SCs) comprise of a large number of exotic phases like 
spin density wave (SDW) \cite{Cruz,Dong,Avci}, orbital order \cite{Shimojima,Chen,sust,Ghosh}, 
structural transition \cite{Sasmal,Huang}, 
nematic order \cite{Baek,Fernandes,arxiv} {\it etc.} These phases are 
sensitive to temperature, pressure, doping concentration 
as well as internal crystallographic structures of the materials
\cite{Mani,Park,sust,Sefat,Jasper}. 
In order to perceive insight on the origin of these intriguing features, enormous efforts  
have already been made by researchers 
through their experimental and theoretical investigations on electronic 
structures of these materials \cite{Kontani,Mazin,Chubukov1,Onari,Stewart,zAs,Jalcom,vca}. 
Among Fe based SCs, 122 family attribute to a large number of anomalies in structural, 
magnetic and superconducting properties including electronic topological transition (ETT) or 
Lifshitz transition \cite{Lifshitz,Cruz,Acta,Liu,Khan}. Unlike oxypnictides, 
high quality large single crystals of 122 Fe-based materials 
can be easily grown by the flux method. One of the popular methods for synthesizing 122 single crystals 
is tin-flux method. Single crystals grown by this tin-flux method, contain unwanted Sn impurity. A sufficient number of 
literatures is available raising the issue of Sn contamination in 122 single crystals. Those investigations 
broach the point of irrefutable alteration in the physical properties of 122 systems in presence of Sn impurity 
by several experiments \cite{Fink, Ni, Su}. Absence of any fruitful and conclusive theoretical study regarding 
the effect of Sn contamination in the 122 systems, make things harder to understand the observed 
anomalies in various physical properties. On the other hand, theoretical study of Lifshitz transition
is of fundamental importance considering its influence in many emergent branches of 
condensed matter physics {\it e.g.,} in topological Dirac semi metal \cite{Xu}, bilayer graphene \cite{Lemonik}, 
quantum Hall liquids \cite{Varlet}, high T$_c$ cuprates \cite{Boeuf} {\it etc.} (Note, the later systems are known to be strongly correlated electron systems whereas the 122 pnictides are considered as weakly correlated one but showing similar effects). In general, the Lifshitz transition is found in some of the 122 family compounds that are grown by various experimental (other than Sn flux) methods \cite{extra}. However, besides the hole or electron doping effects, the localization effects from impurity scattering may significantly affect the Fermi surface as well \cite{extra1} and the contamination effect by impurity is not restricted to Sn-flux method only (see for example occurence of Lifshitz transition as an effect of Ca impurities being unintentionally present in  fcc-Yb crystals \cite{extra2}).  Therefore, our study is of general significance.   \\

Single crystals of BaFe$_2$As$_2$ (Ba122), SrFe$_2$As$_2$ (Sr122) and CaFe$_2$As$_2$ (Ca122), synthesized from 
tin-flux method contain a few percentage (1\%-5\%) of 
Sn incorporated into the crystal structure. 
As a result of Sn impurity, the crystallographic structure of 122 systems get modified \cite{Su,Rotter,thesis}.
It was found that 95\% of the 2a site (0,0,0) is occupied by Ba atoms and the rest are found to be replaced by 
Sn atoms at the site 4e (0,0,$z$ with $z$=$0.093$) \cite{thesis}. 
It is obvious from current literature that electronic structures of 122 systems 
are highly influenced by certain moderation of structural parameters \cite{zAs,Acta,pla}. 
This incorporation of Sn impurity leads to 
certain changes in the physical properties of Ba122 system. The presence of a small fraction 
of Sn impurity in BaFe$_2$As$_2$ (Ba122) system 
curtails down the structural as well as SDW transitions from 138K to 85K and also give rise to a notable difference in 
the thermal behaviour of electrical resistivity and magnetic susceptibility \cite{Su,Wang,Colombier,Kim}. On the other 
hand, no significant modifications in the physical properties 
of Sr122 and Ca122 systems are observed due to the presence of Sn impurity. Since these structural 
and magnetic transitions are intimately related to the electronic structures, it calls for the study 
of detailed electronic structure through 
density of states (DOS), band structures and Fermi surfaces of BaFe$_2$As$_2$ in presence of Sn impurity.
Such study is absent from current literature.
Within virtual crystal approximation, we have made the first attempt to explain 
the observed diversities in the physical properties of these 122 materials due to Sn impurity. 
Our theoretical investigation using first principles density functional theory 
reveals that in Ba122 system, an impurity induced Lifshitz 
transition occurs but such transition is absent in other 122 systems like Sr122 and Ca122 systems. 
This may be the reason for observed differences in the physical properties of various 122 materials.
Our manifestation of Sn impurity in the crystal structures through VCA method, slightly 
deviates from the ideal experimental conditions. We discuss this issue in the theoretical method section 
along with the details of calculations. In the next section, we present our detail calculated 
DOS, band structures and FSs of Ba122, Sr122 and Ca122 systems with inbuilt Sn impurity. In the last 
section we present the summary of our work with conclusions. 
\section{Theoretical method}
Our first principles electronic structure calculations are performed by using CASTEP which is a
plane wave psudopotential method based on density functional theory \cite{CASTEP},
where the electronic exchange correlation is treated under the generalized gradient approximation 
(GGA) using Perdew-Burke-Ernzerhof (PBE) functional \cite{PBE}. Experimentally measured orthorhombic (20K) 
as well as tetragonal (300K) structural parameters {\it i.e.,} 
$a$, $b$, $c$ and $z_{As}$ (fractional z co-ordinate of As atom)\cite{Rotter,thesis,Tegel,Goldman} 
are used as inputs of our first principles 
density of states (DOS), band structures and Fermi surface calculations.
Implementation of Sn impurity in BaFe$_2$As$_2$ and SrFe$_2$As$_2$ systems has been treated within 
virtual crystal approximation. Virtual crystal approximation is an well 
known method for treating disordered systems within primitive unit cell \cite{Bellaiche}.
This VCA method forges us to use the same atomic co-ordinates for Ba and Sn atoms which is not 
the actual condition where Sn atoms are situated at slightly different co-ordinates (0, 0, 0.093) close to Ba (0, 0, 0) atom.
(In general, Sn shows a chemical similarity to both neighbouring group-14 elements, 
germanium and lead, and has two possible oxidation states, +2 and slightly more stable +4.)
This displacement of Sn atom however, is accounted for the stereo-chemical active lone pair of the 
formal divalent Sn atom that results in a change of the coordination sphere which is a Sn$^{2+}-$typical square 
pyramidal coordination by four As atoms \cite{thesis}.
In this work, we use the virtual crystal approximation (VCA) method, developed by Bellaiche 
and Vanderbilt \cite{Bellaiche} based on weighted averaging of pseudopotentials. In VCA, 
a doped crystal with original periodicity but composed of fictitious  `virtual atoms' is 
created to mimic the actual doped system. For example, one can construct a virtual atom 
like A$_{1-x}$B$_x$ for a single lattice site, where $x$ is the doping concentration and thus overlooks any possible short range order.
One uses ultra-soft pseudopotentials which are prone to generating ghost 
states in the VCA framework, while none of the individual potentials has a ghost state. 
There is a standard way of testing the VCA approach for ghost states, so one should always 
examine the value of the derivative of the total energy w.r.t. E$_{cut}$ (E$_{cut}$  
defines the size of plane wave basis set) during the finite basis set correction calculation. 
It should be of the same order as the derivatives for the end-member structures \cite{link}. 
We have checked this value for all these three systems 
up to 5$\%$ Sn doping in order to inspect the applicability of VCA method in these cases.
Both spin-polarized (for orthorhombic phase with space group symmetry Fmmm [No. 69]) as well as
non-spin-polarized single point energy calculations 
are carried out (for tetragonal phase with space group symmetry I4/mmm [No. 139]) 
using ultrasoft pseudo-potentials and plane-wave basis 
set with energy cut off 600 eV and self-consistent field (SCF) tolerance $10^{-6}$ eV/atom. 
Brillouin zone is sampled in the k space within Monkhorst–Pack scheme
and grid size for SCF calculation is chosen as $26\times 26\times 26$. 
\section{Results and discussion}
Our main aim is to examine effect of Sn impurity in the electronic structures of Ba122, Sr122 and Ca122 systems within VCA method. 
We calculate DOS, band structures and FSs of these three systems (pure and Sn contaminated) using experimental 
lattice parameters ($a$, $b$, $c$, z$_{As}$) in the orthorhombic (low temperature) phases. 
No significant differences in the calculated DOS for pure and Sn contaminated 122 systems are observed. 
\begin{figure}
 \centering
 \includegraphics[height=11cm,width=11cm]{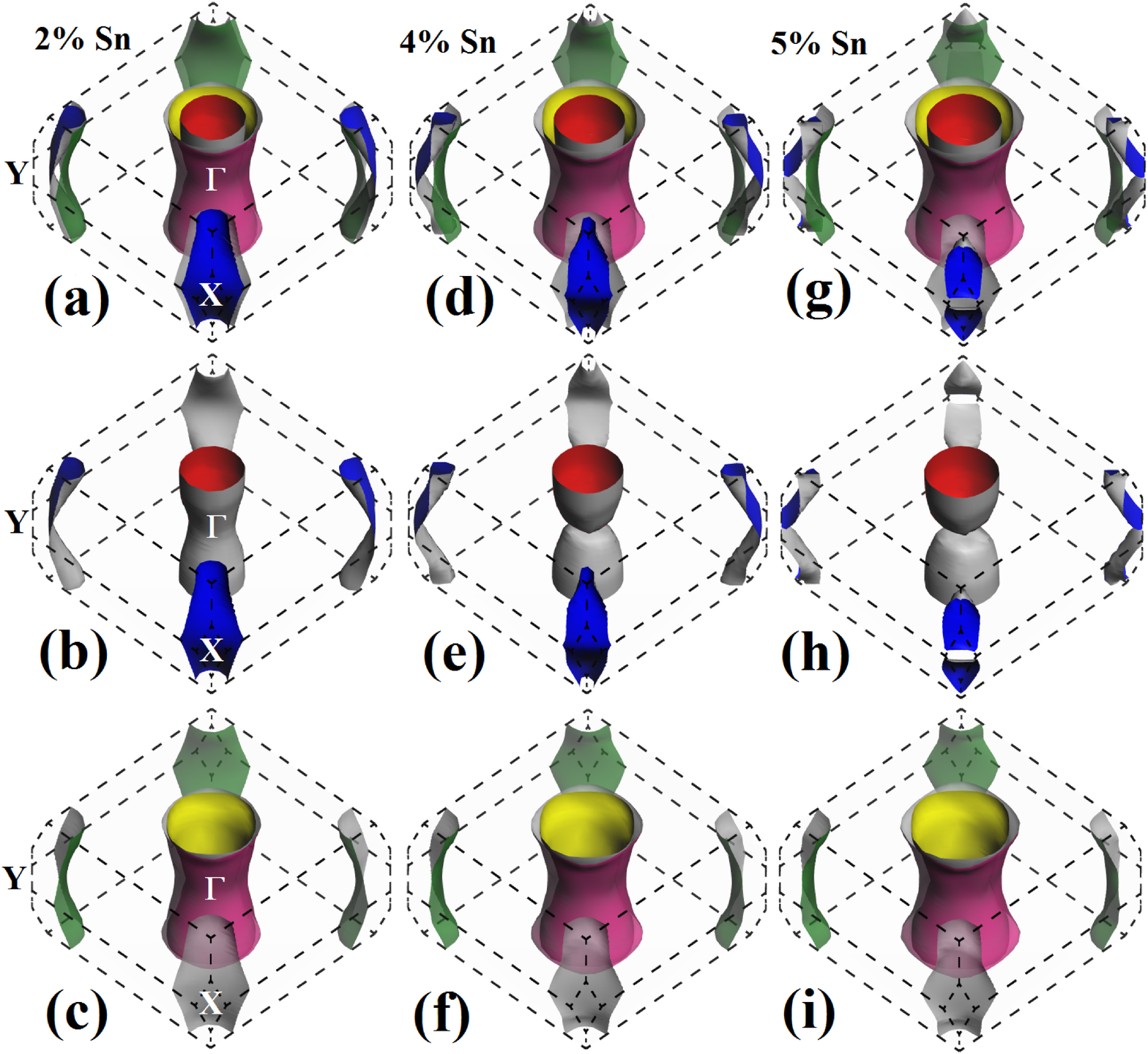}
 \caption{Calculated Fermi surfaces in orthorhombic phase within VCA method for (a, b, c) 
 Ba$_{0.98}$Sn$_{0.02}$Fe$_2$As$_2$, (d, e, f) Ba$_{0.96}$Sn$_{0.04}$Fe$_2$As$_2$ and (g, h, i) 
 Ba$_{0.95}$Sn$_{0.05}$Fe$_2$As$_2$ systems. For clarity individual FSs are shown separately 
 in 2nd and 3rd row. Different colours are used to indicate different FSs. 
 k points in the Brillouin Zone (BZ) are also indicated 
 in (a). Shrinkage of electron like FSs and surging of hole like FSs are worth noticing. 
 This indicates hole doping.}
\label{FS-vca}
\end{figure}
\begin{figure}
 \centering
 \includegraphics[height=11cm,width=9cm]{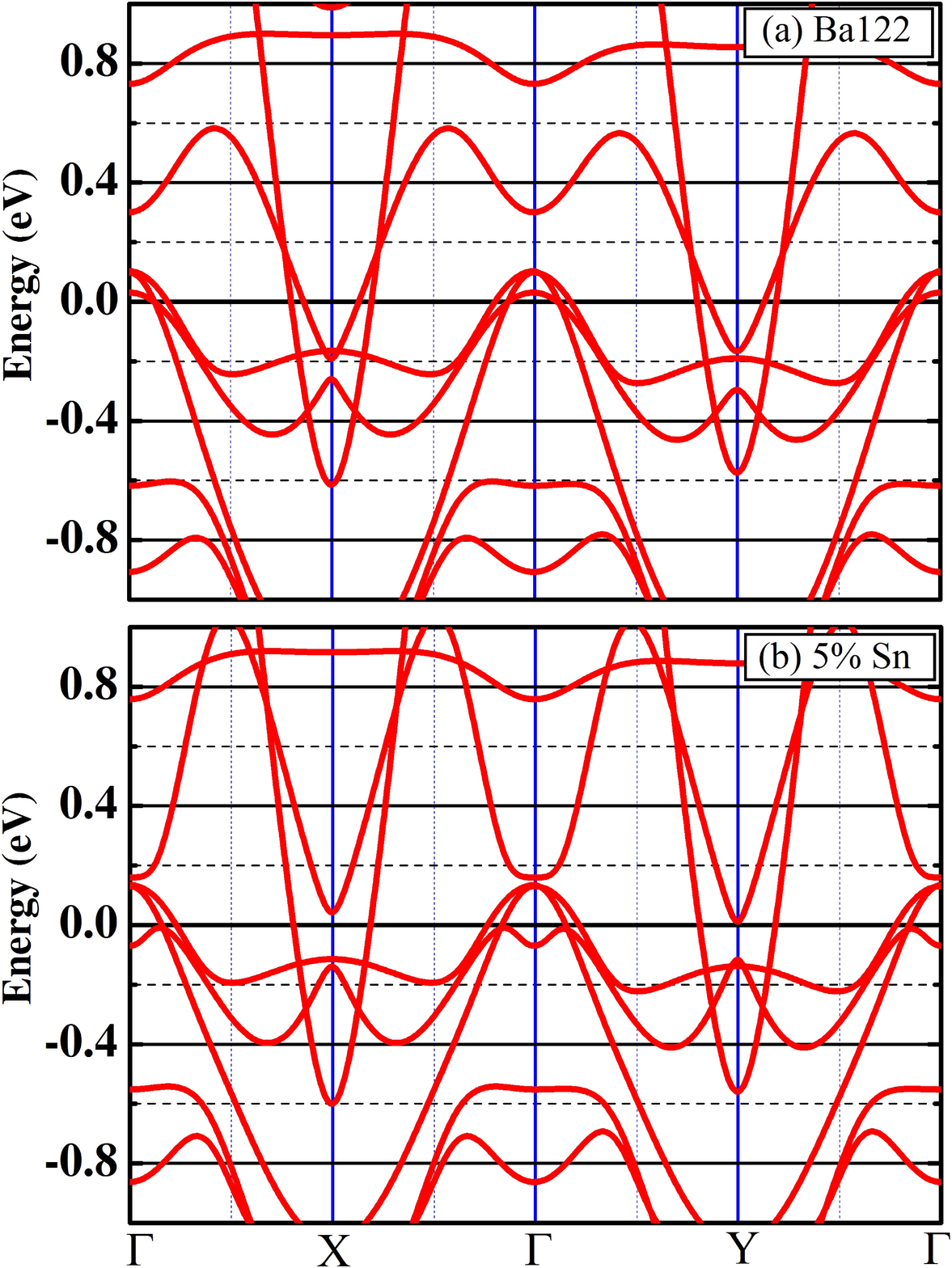}
 \caption{Calculated band structures in orthorhombic phase within VCA method for (a) 
 BaFe$_2$As$_2$ and (b) Ba$_{0.95}$Sn$_{0.05}$Fe$_2$As$_2$ systems along $\Gamma-X-\Gamma-Y-\Gamma$ k-path in the 
 BZ. In comparison to FIG.\ref{BS-vca-Sr-fp}, 5\% Sn doped Ba122 system shows significant modification 
 in band structure as far as electronic topological transitions are concerned.}
\label{BS-vca-Ba-fp}
\end{figure}
\begin{figure}
 \centering
 \includegraphics[height=11cm,width=11cm]{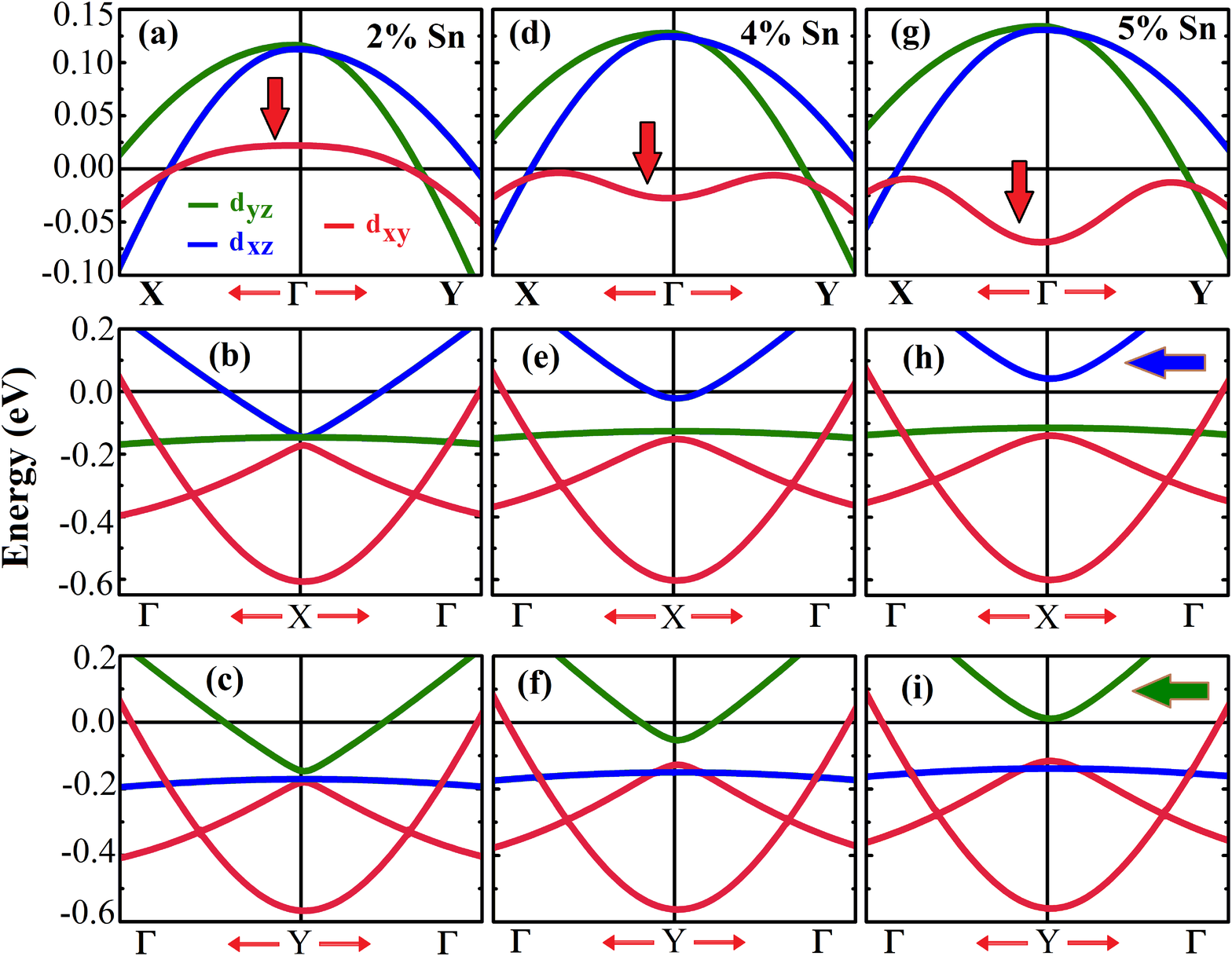}
 \caption{Calculated band structures in orthorhombic phase within VCA method for (a, b, c) 
  Ba$_{0.98}$Sn$_{0.02}$Fe$_2$As$_2$, (d, e, f) Ba$_{0.96}$Sn$_{0.04}$Fe$_2$As$_2$ and (g, h, i) 
  Ba$_{0.95}$Sn$_{0.05}$Fe$_2$As$_2$ systems around $\Gamma$ (1st row), X (2nd row) and Y (3rd row) points. 
  Electronic topological transition (transition of hole like band to electron like band) is indicated 
  using red arrow around $\Gamma$ point. Movement of the d$_{xz}$ (d$_{yz}$) band around X (Y) point 
  away from the Fermi level is shown by blue (green) arrow. This figure may be contrasted with FIG.\ref{BS-vca-Sr}
   and FIG.\ref{BS-vca-Ca}}
\label{BS-vca}
\end{figure}
\begin{figure}
 \centering
 \includegraphics[height=11cm,width=8cm]{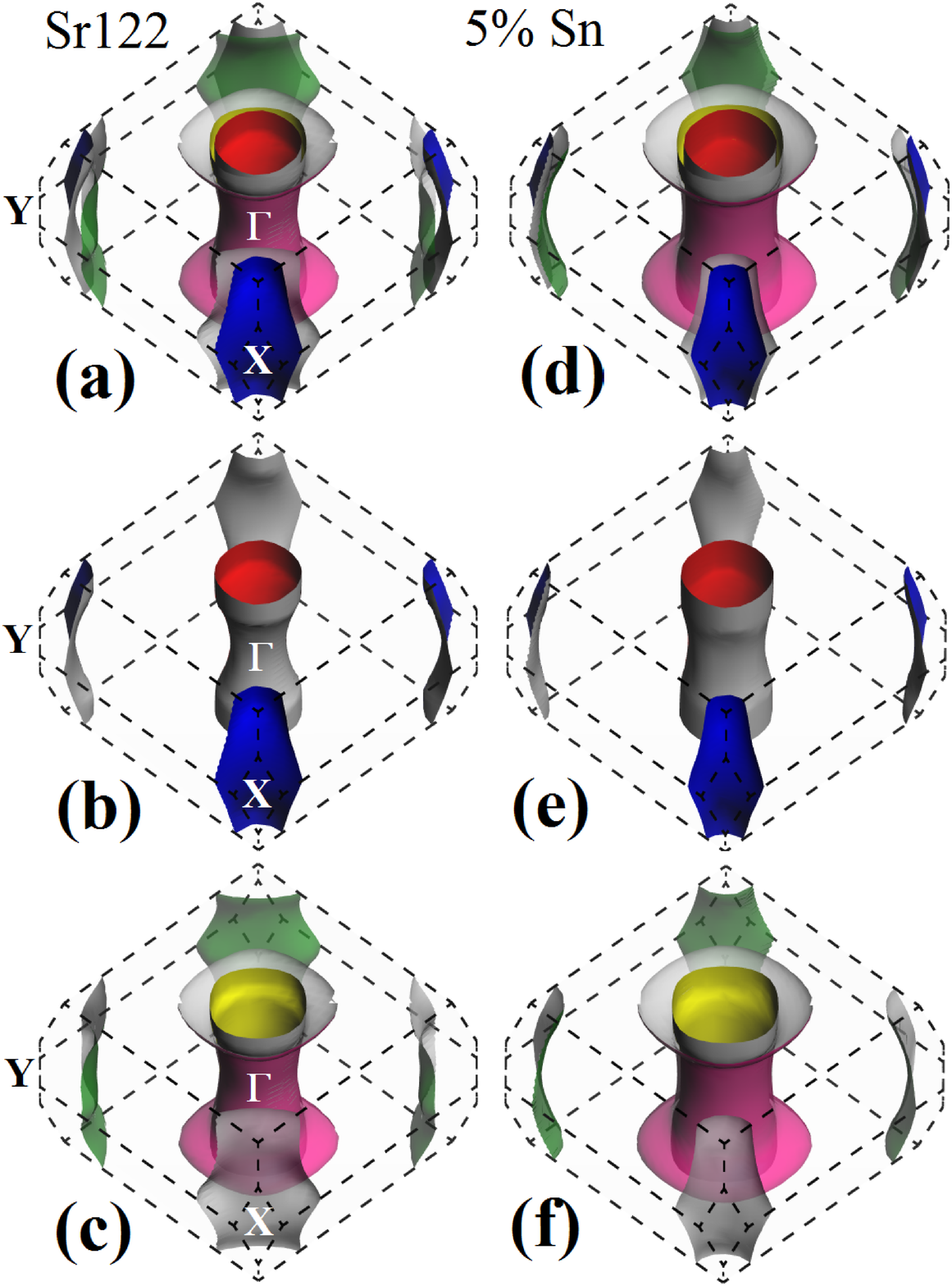}
 \caption{Calculated Fermi surfaces in orthorhombic phase within VCA method for (a, b, c) 
  SrFe$_2$As$_2$ and (d, e, f) Sr$_{0.95}$Sn$_{0.05}$Fe$_2$As$_2$ systems. 
  For clarity individual FSs are shown separately 
  in 2nd and 3rd row. Shrinkage of electron like FSs and surging of hole like FSs are worth noticing. 
  This indicates hole doping.}
\label{FS-vca-Sr}
\end{figure}
We exhibit our calculated electronic structure in the orthorhombic phase  
where experimentally measured orthorhombic lattice parameters are used as the input of our first 
principles calculations. It should be mentioned here that the structural transition in these 
materials is due to the orbital ordering between d$_{yz}$ and d$_{xz}$ bands around high symmetry k points \cite{sust}. 
In FIG.\ref{FS-vca} 
we depict the FSs of BaFe$_2$As$_2$ systems with various percentages of Sn impurity. 
There are two electron like FSs around the four corners and three hole like FSs at the centre of the Brillouin Zone (BZ). 
In order to provide a clearer view of all the FSs, we present the FSs separately in second and third rows 
of FIG.\ref{FS-vca}. One can see from these FSs that with increasing Sn impurity electron like FSs 
shrink and hole like FSs expand, which corroborates with the previous experimental picture of hole doping 
as a consequence of Sn impurity in Ba122 structure \cite{Colombier}. One of our most important observations, is the occurrence of 
Lifshitz transition upon certain percentage of Sn impurity (see FIG.\ref{FS-vca} and FIG. \ref{BS-vca-Ba-fp}).
With 4\% Sn impurity in Ba122 system, the innermost FS at the centre of the BZ vanishes around $\Gamma$ point 
as a direct consequence of Lifshitz transition.
\begin{figure}
 \centering
 \includegraphics[height=11cm,width=9cm]{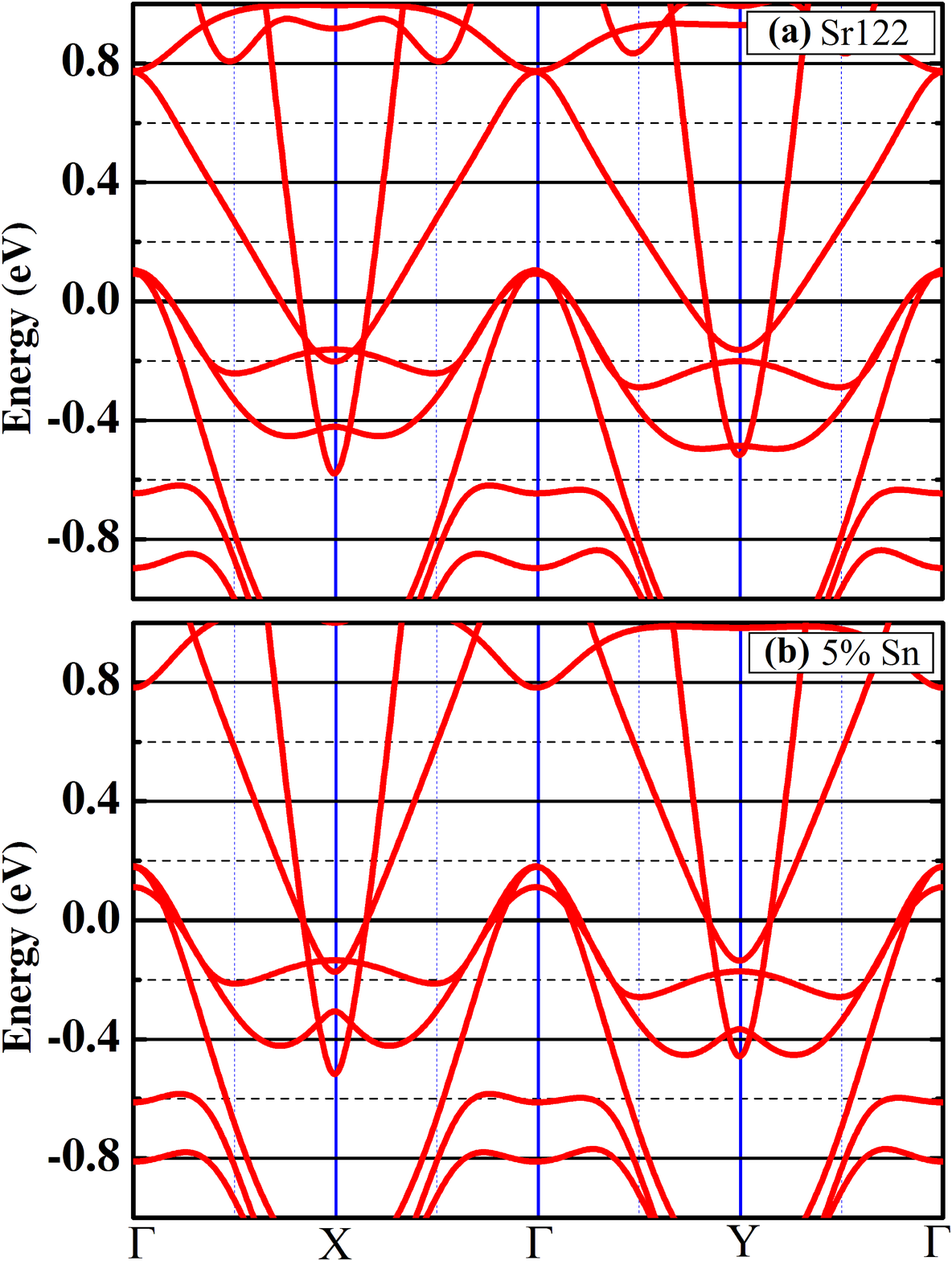}
 \caption{Calculated band structures in orthorhombic phase within VCA method for (a) 
 SrFe$_2$As$_2$ and (b) Sr$_{0.95}$Sn$_{0.05}$Fe$_2$As$_2$ systems along $\Gamma-X-\Gamma-Y-\Gamma$ k-path in the 
 BZ. In comparison to FIG.\ref{BS-vca-Ba-fp}, 5\% Sn doped Sr122 system shows insignificant modification 
 in band structure as far as electronic topological transitions are concerned.}
\label{BS-vca-Sr-fp}
\end{figure}

On the whole with increasing Sn percentage, the shapes of FSs become more 3D like, resulting  
in large degradation of FS nesting. As a result of degradation of nesting 
magnetic order becomes weaker which might be the reason of lowering of structural as well as SDW transition 
temperatures (T$_s$ and T$_{SDW}$) in Sn contaminated Ba122 systems. To illustrate the observed Lifshitz transition, 
we exhibit the calculated band structures of Sn contaminated Ba122 system for 2\%, 4\% and 5\% Sn impurity in FIG.\ref{BS-vca}. 
Band structures around $\Gamma$, X and Y points are shown in first, second and third rows of FIG.\ref{BS-vca} 
respectively. It is quite evident from these figures that at around 4\% Sn impurity Lifshitz transition occur 
in Ba122 system. As a signature of this Lifshitz transition, a hole like band (d$_{xy}$ band around $\Gamma$ point) 
becomes electron like (see FIG.\ref{BS-vca}g). 
Observation of Lifshitz transition in other high symmetry k points also occur as clearly 
demonstrated in FIG.\ref{BS-vca}.
This type of transition can create remarkable topological modification of FSs, 
which might be the reason of alteration of superconducting properties of these Ba122 systems with Sn impurity. 
Transition of hole like band to electron like band as well as disappearance of electron 
like band from Fermi level (FL) would not only modify the shape of the FSs but also affect the nesting 
condition between electron and hole like bands severely. Since nesting of FS is directly related to 
magnetic SDW transition, lowering of observed magnetic transition in Sn contaminated 
Ba122 system may be attributed to the electronic topological transition as 
found from our first principles simulations.
Band structures around X and Y points reveal that orbital order also got modified in the presence of 
slightly higher percentages of Sn impurities ($>$1\%). This might stand out as an explanation of reduction of 
structural transition temperature (T$_s$). On the other hand, there is no experimental evidence of lowering 
structural transition temperature (T$_s$) and SDW transition temperature (T$_{SDW}$) in Sr122 and Ca122 
compounds, prepared by Sn-flux method (containing Sn impurity). We have also performed electronic 
structure calculation of Sn contaminated Sr122 and Ca122 systems. In FIG.\ref{FS-vca-Sr}  
we have displayed the FSs of Sr122 system (first column) as well as 5\% Sn contaminated Sr122 system 
(second column). In FIG.\ref{FS-vca-Ca} we present the FSs of Ca122 system (first column) 
as well as 5\% Sn contaminated Ca122 system 
(second column). Calculated FSs of Sr122 and Ca122 system with 5\% Sn impurity (FIG.\ref{FS-vca-Sr} and FIG.\ref{FS-vca-Ca}) 
give a clear indication of hole doping {\it i.e.,} expansion of hole FSs and shrinkage of electron FSs. 
However, there is no trace of Lifshitz transition in case of Sn contaminated 
Sr122 and Ca122 systems contrary to the case of Ba122 system. Besides there is no 
significant change in the FS topology specially, the change in dimensionality of FSs in presence of 5\% Sn impurity. 
So there is no evidence of degradation of FS nesting in these two systems. 
This may also stand out as an explanation for the robustness of magnetic, structural and superconducting transition 
temperatures for Sr122 and Ca122 systems with Sn impurity. 
\begin{figure}
 \centering
 \includegraphics[height=11cm,width=8cm]{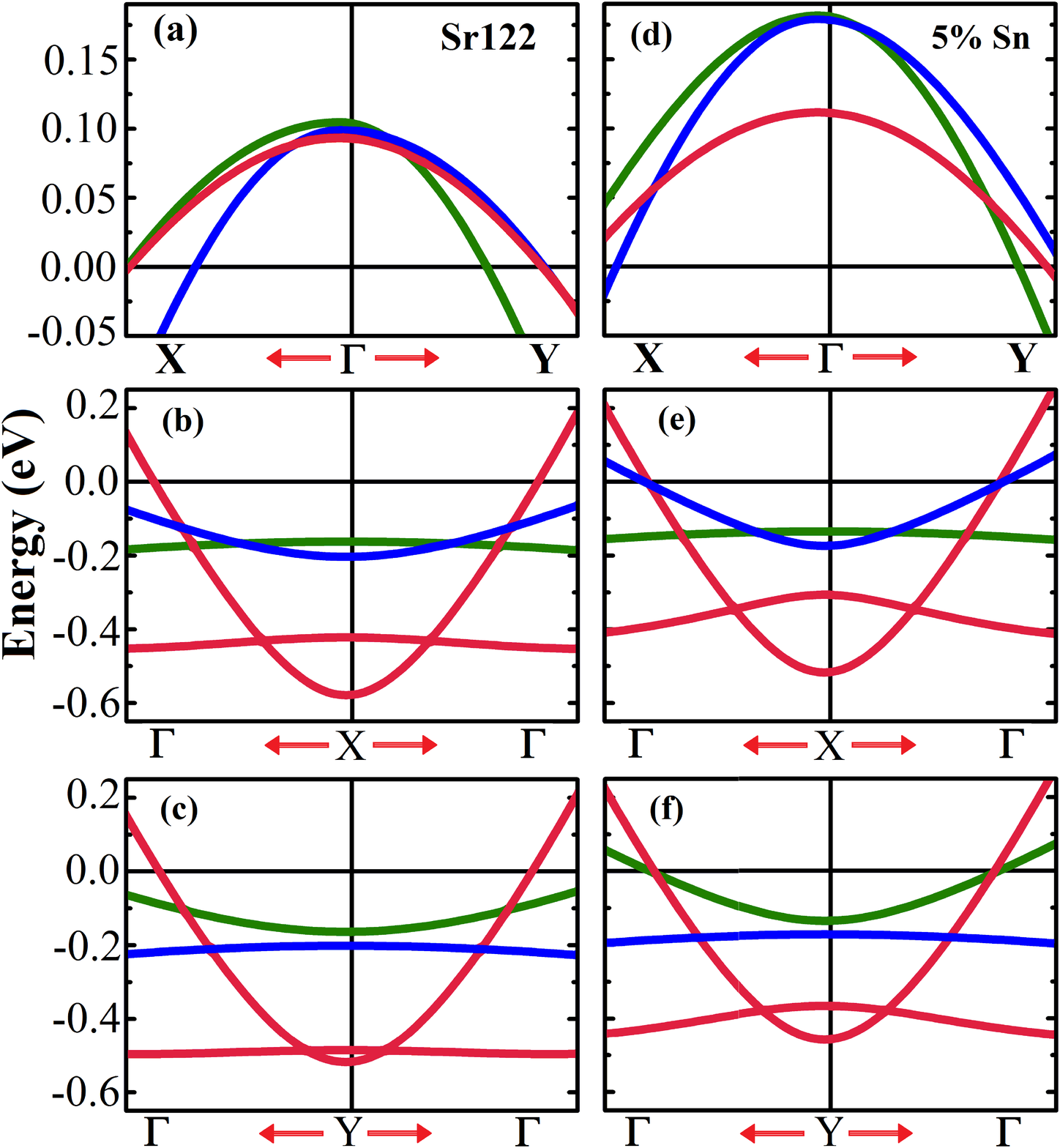}
 \caption{Calculated band structures in orthorhombic phase within VCA method for (a, b, c) 
  SrFe$_2$As$_2$ and (d, e, f) Sr$_{0.95}$Sn$_{0.05}$Fe$_2$As$_2$ systems around $\Gamma$ (1st row), 
  X (2nd row) and Y (3rd row) points. 
  Modification in the band widths of electron (reduction in band width) and 
  hole (increment in band width) like bands causes apprehendable modifications in the FSs (See FIG.\ref{FS-vca-Sr}).}
\label{BS-vca-Sr}
\end{figure}
We have also simulated the band structure of Sr122 and Ca122 systems in order to see the moderation of 
orbital order if any. In FIG.\ref{BS-vca-Sr-fp}a, the band structure 
along k path $\Gamma-X-\Gamma-Y$ of pure Sr122 system
and in FIG.\ref{BS-vca-Sr-fp}b that for 5\% Sn contaminated Sr122 systems are presented. 
Similarly in FIG.\ref{BS-vca-Ca-fp}a, the band structure 
along k path $\Gamma-X-\Gamma-Y$ of pure Ca122 system
and in FIG.\ref{BS-vca-Ca-fp}b that for 5\% Sn contaminated Ca122 systems are depicted. 
Band structures of these pure and impure systems have no qualitative differences. 
In FIG.\ref{BS-vca-Sr} and FIG.\ref{BS-vca-Ca} band structure around $\Gamma$, X and Y points 
of pure as well as 5\% Sn contaminated Sr122 and Ca122 systems have been shown. There is no sign of Lifshitz 
transition or any kind of electronic topological transition (ETT) at least up to 5\% Sn impurity. 
And there is almost no change in the nature of orbital ordering between d$_{yz}$ and d$_{xz}$ bands around 
$\Gamma$, X and Y point with the introduction of Sn impurity (5\% Sn) in Sr122 and Ca122 systems.
Since orbital ordering of d$_{yz}$ and d$_{xz}$ bands around X and Y points are the major contributors 
to the structural transition from tetragonal (high temperature) to orthorhombic phase (low temperature) 
in 122 systems \cite{arxiv,sust}, no significant moderation of orbital 
ordering around X and Y points support the experimental observation of 
the unaffected structural transition temperature in Sn induced Sr122 and Ca122 systems.
Lifshitz transition with variation of pressure or doping concentration or even with variation of temperature, 
plays an important role in 122 Fe-based superconductors and is observed 
experimentally \cite{Liu,Khan}. Thus one must be sure about the 
fact whether the observed Lifshitz transitions are artefacts of Sn impurity or not. 
So by using VCA method for introducing Sn into the Ba122/Sr122/Ca122 structure, we provide a detailed 
electronic structures for pure and Sn contaminated systems for the first time. We also elucidate the experimentally 
observed anomalies for these Sn induced 122 systems through our first principles electronic structure 
calculation which is not available in the current literature.
\begin{figure}
 \centering
 \includegraphics[height=11cm,width=8cm]{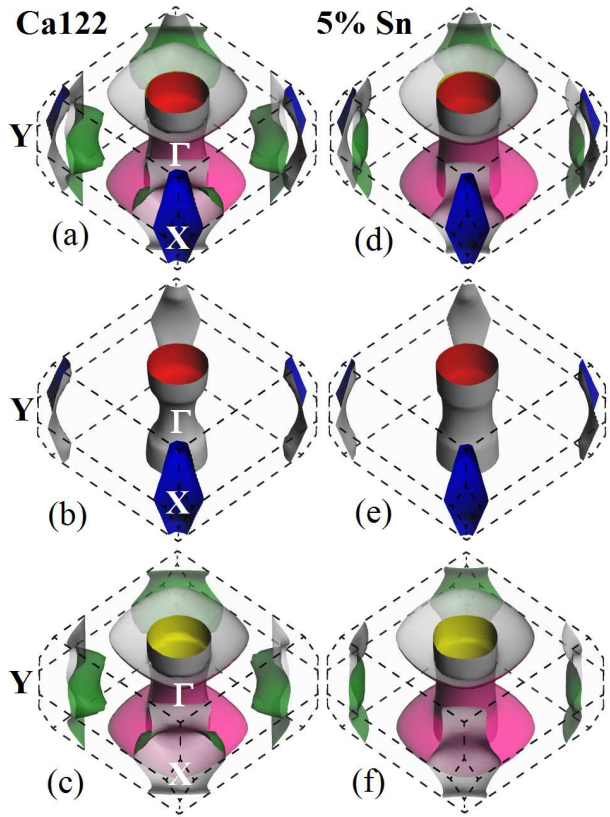}
 \caption{Calculated Fermi surfaces in orthorhombic phase within VCA method for (a, b, c) 
   CaFe$_2$As$_2$ and (d, e, f) Ca$_{0.95}$Sn$_{0.05}$Fe$_2$As$_2$ systems. 
   For clarity individual FSs are shown separately 
   in 2nd and 3rd row. Shrinkage of electron like FSs and surging of hole like FSs are worth noticing. 
   This indicates hole doping.}
\label{FS-vca-Ca}
\end{figure}

In order to understand the influence of Sn doping on various 122 materials it is essential to identify 
that the band structures of undoped Ba122, Sr122 and Ca122 systems (in the orthorhombic phase) 
already have significant differences.
 \begin{figure}
  \centering
  \includegraphics[height=11cm,width=9cm]{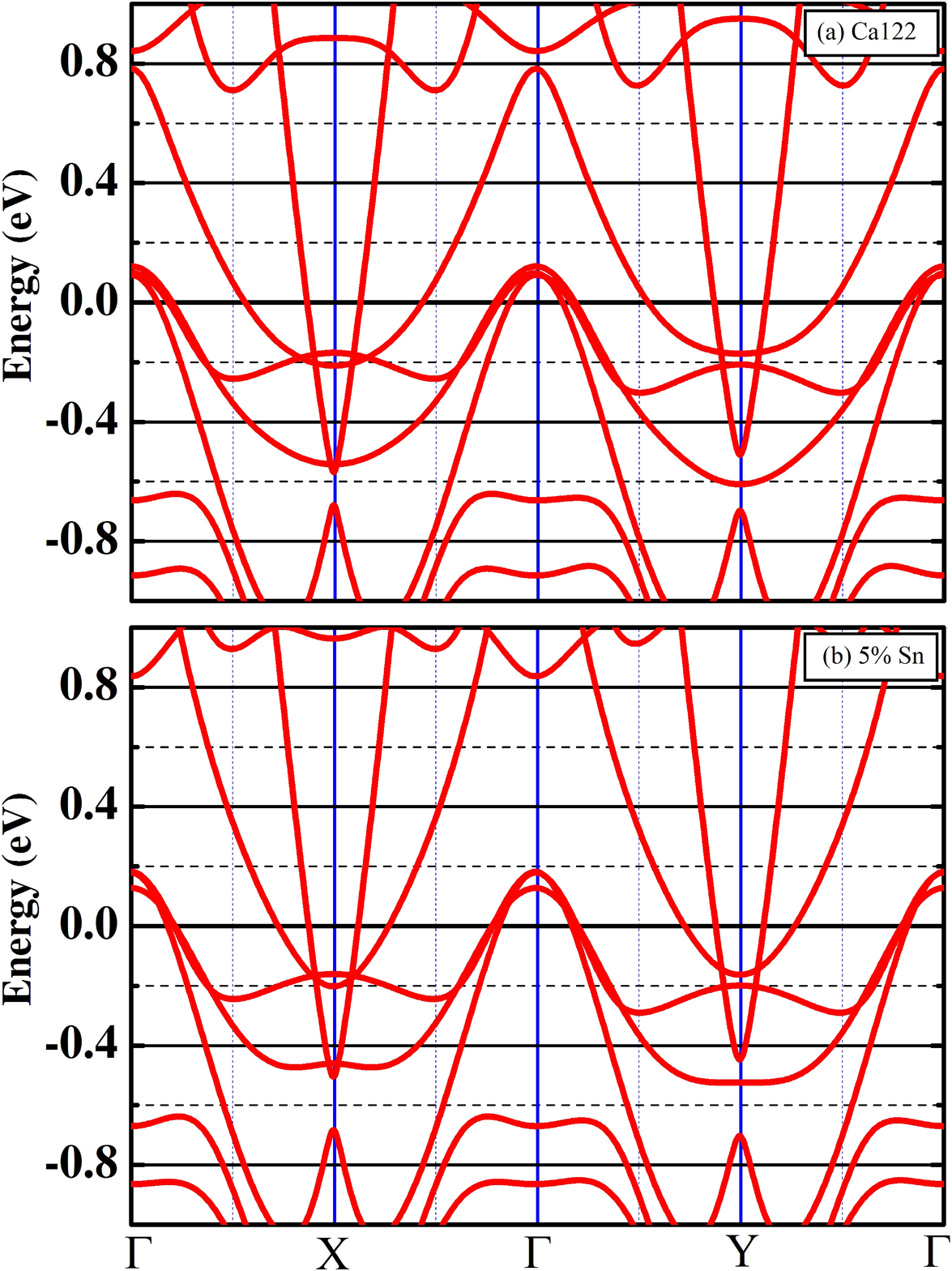}
  \caption{Calculated band structures in orthorhombic phase within VCA method for (a) 
  CaFe$_2$As$_2$ and (b) Ca$_{0.95}$Sn$_{0.05}$Fe$_2$As$_2$ systems along $\Gamma-X-\Gamma-Y-\Gamma$ k-path in the 
  BZ. In comparison to FIG.\ref{BS-vca-Ba-fp}, 5\% Sn doped Ca122 system shows insignificant modification 
  in band structure as far as electronic topological transitions are concerned.}
 \label{BS-vca-Ca-fp}
 \end{figure} 
For Ba122 system, electronic bands around $\Gamma$ point is very close to the Fermi level (FL), specially, 
the flat d$_{xy}$ band nearly touches the FL. The other two bands d$_{xz}$ and d$_{yz}$ are slightly 
higher in energy compared to that of the d$_{xy}$. This should be contrasted with that of the Sr122 and Ca122 systems 
(see FIG.\ref{BS-vca-Ba-fp}, \ref{BS-vca-Sr-fp}, \ref{BS-vca-Ca-fp}), where all the d$_{xy}$, d$_{xz}$ and d$_{yz}$) bands 
are further away from FL. This is because of the largest size of Ba (253 pm) compared to that of 
Sr (219 pm) and Ca (194 pm) causing two Fe-As layers furthest in Ba122 compared to Sr122, Ca122. As a result Ba-122 is more planer
compared to others (see for example, a dimensional crossover of FS in 122 systems \cite{pla} )resulting a flatter d$_{xy}$ band near FL 
It is to be noted that the d$_{xy}$ band 
 is  planar ($xy$) in nature in contrast to 
the d$_{xz}$/d$_{yz}$ bands. Sn substitution in the M-atoms (Ba/Sr/Ca) 
is an out-of-plane substitution. Ba atom being the largest in size among the 
M-atoms as well as Sn, a few percentage of Sn substitution keeps the 
effective size of the  `virtual M atom' largest in size in case of Ba, smaller in case of Sr
 and the smallest in case of Ca. This would cause vertical movement between the Fe-As planes 
 least for Sn-doped Ba-122 but larger for Sn-doped Sr-122 and largest for Sn-doped Ca-122. 
 Therefore, naturally d$_{xz}$/d$_{yz}$ bands would move furthest in case 
 of Sn-doped Ca-122 whereas least movement to the d$_{xy}$ bands (see for example, Figures 3, 6, 9). 
 That also simultaneously explains hole doping effect.
 \begin{figure}
  \centering
  \includegraphics[height=11cm,width=8cm]{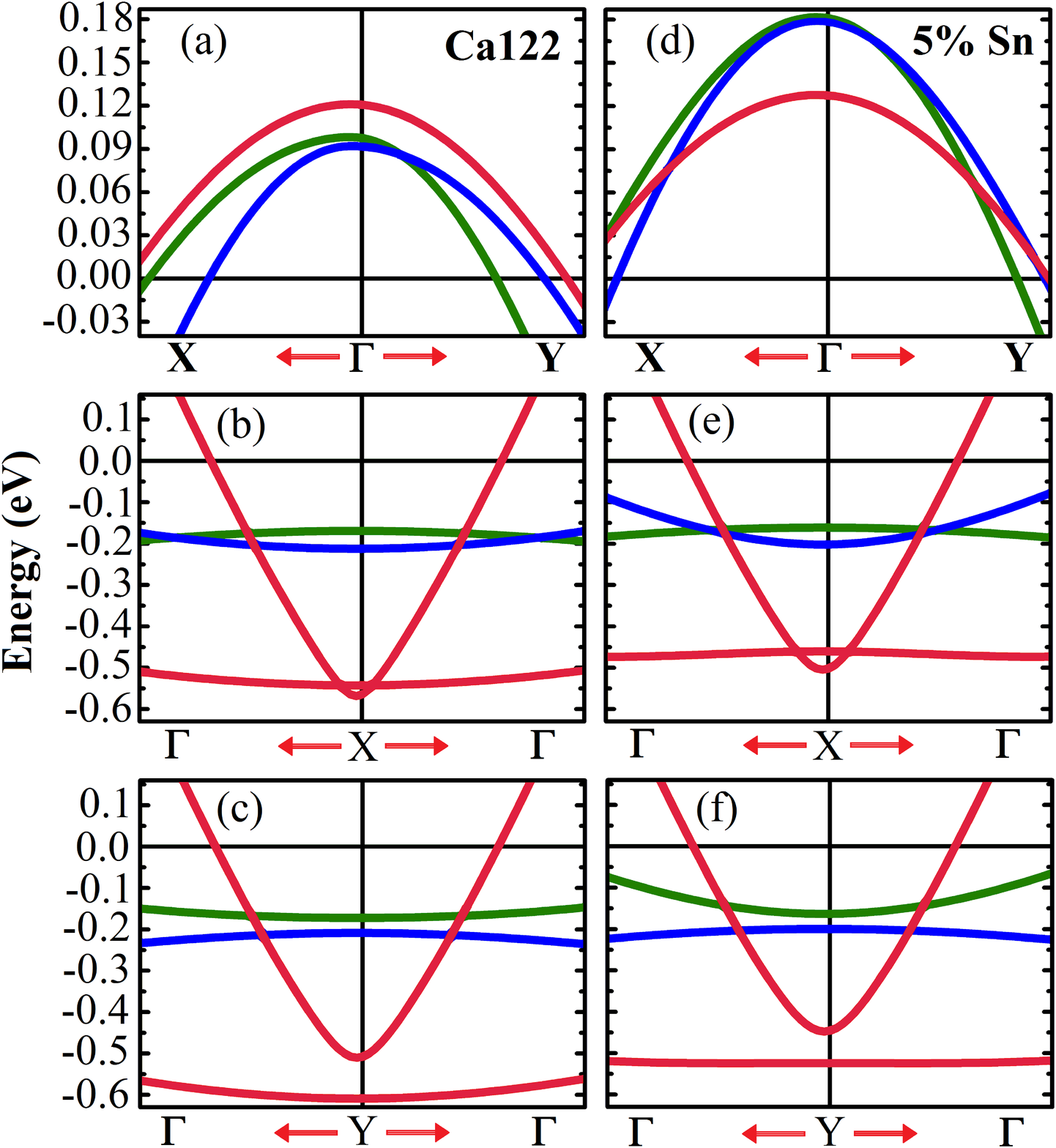}
  \caption{Calculated band structures in orthorhombic phase within VCA method for (a, b, c) 
    CaFe$_2$As$_2$ and (d, e, f) Ca$_{0.95}$Sn$_{0.05}$Fe$_2$As$_2$ systems around $\Gamma$ (1st row), 
    X (2nd row) and Y (3rd row) points. Modification in the band widths of electron (reduction in band width) and 
   hole (increment in band width) like bands causes apprehendable modifications in the FSs (See FIG.\ref{FS-vca-Ca}).}
 \label{BS-vca-Ca}
 \end{figure}
 Thus, Sn impurity causes splitting between the d$_{xy}$ and the pair of \lq nearly \rq 
 degenerate bands d$_{xz}$/d$_{yz}$. The modulations of the d$_{xz}$/d$_{yz}$ bands 
 can also produce modulations in the d$_{xy}$ bands. But the  d$_{xy}$ band around the $\Gamma$ 
 point is much lower in energy compared to that of the d$_{xz}$/d$_{yz}$ bands and 
 closer to the FL for the undoped Ba122. This is not the case of undoped Sr122, Ca122 
 where d$_{xy}$/d$_{xz}$/d$_{yz}$ are nearly at same but higher in energy. The effect 
 of modulations in the d$_{xy}$ bands due to that of the d$_{xz}$/d$_{yz}$ bands is thus 
 minimum in case of Sr-122, Ca-122 systems. The consequence of Sn impurity on 122 family materials 
 thus depend on strong doping sensitivity of the underlying system --- energy levels of 
 d$_{xy}$/d$_{xz}$/d$_{yz}$ and the impurity energy level. The d$_{xy}$ band of Ba122 being already very 
 close to the FL in presence of Sn impurity is pushed down below the FL due to such modulation, 
 causing the Lifshitz transition.

\section{Conclusion}

We have studied the effect of Sn impurity on the electronic structures of 122 systems 
(BaFe$_2$As$_2$, SrFe$_2$As$_2$ and CaFe$_2$As$_2$) using VCA method 
for introducing Sn impurity into the 122 crystal structures.
Presence of Sn impurity modifies the electronic structures of Ba122 system 
remarkably, which in turn is expected to
result certain changes in the physical properties like reduction of structural and magnetic transition temperatures 
as well as superconducting transition temperature. On the contrary, Sn impurity does not have much impact on the 
electronic structures of Sr122 and Ca122 systems as reflected in the experimental evidence of robustness of structural, 
magnetic and superconducting transition temperatures \cite{Su,Sun}. In Ba122 system, 
we observe Sn impurity induced Lifshitz transition 
around $\Gamma$ point that dictates several changes in the electronic 
structures near Fermi level including degradation of FS nesting contrasting the case of 
Sr122 and Ca122 systems. Orbital ordering between 
d$_{yz}$ and d$_{xz}$ bands around X and Y points, which is responsible for structural transition, 
remains unaltered in case of Sr122 and Ca122 systems but is substantially modified in case of 
Ba122 system in presence of Sn impurity. 
This work thus not only provide a detailed description of electronic structures of Sn contaminated 122 systems 
but also provides a possible explanation to the various observed anomalies in the physical properties 
that are not well understood so far. We believe our work will lead to further similar research beyond VCA. 

\section{Acknowledgments}

One of us (SS) acknowledges the HBNI, RRCAT for financial support and encouragements. 
We thank Dr. P. A. Naik and Dr. P. D. Gupta for their encouragement in this work.

\end{document}